\documentstyle[amssymb,prl,aps,twocolumn,epsf,floats]{revtex}

\newcommand{\ie}{\mbox{i.\ e.\ }}
\newcommand{\eg}{\mbox{e.\ g.\ }}

\newcommand{\etc}{\mbox{etc.}}
\newcommand{\etal}{\mbox{\it et.\ al.\ }}
\newcommand{\be}{\begin{eqnarray}}
\newcommand{\en}{\end{eqnarray}}
\newcommand{\no}{\nonumber}

\newcommand{\mc}{\mathcal}
\newcommand{\up}{c_{\uparrow}}
\newcommand{\dn}{c_{\downarrow}}

\newcommand{\f}{f}
\newcommand{\p}{p}
\newcommand{\fd}{f^{\dagger}}
\newcommand{\pd}{p^{\dagger}}
\newcommand{\at}{\tilde{a}}
\newcommand{\atd}{\tilde{a}^{\dagger}}

\begin{document}

\twocolumn[\hsize\textwidth\columnwidth\hsize\csname
@twocolumnfalse\endcsname

\title{Order from disorder in the double--exchange model}

\author{Nic Shannon}
\address{
Max--Planck--Institut f{\"u}r Physik komplexer Systeme,
N{\"o}thnitzer Str. 38, 01187 Dresden, Germany.
}
\author{Andrey V.\ Chubukov}
\address{
Department of Physics, University of Wisconsin--Madison,
1150 Univ. Av., Madison WI 53706, USA.
}
\date{\today}
\maketitle
\begin{abstract}
The magnetic excitations of the double exchange (DE) model are 
usually discussed in terms of an equivalent ferromagnetic Heisenberg 
model.
However this equivalence is valid {\it only} at a quasi--classical
level --- we show that both quantum and thermal corrections to the 
magnetic 
properties of DE model differ from {\it any} effective Heisenberg 
model
because its spin excitations interact only indirectly, through the 
exchange 
of charge fluctuations.    
We also find that the  competition between ferromagnetic double 
exchange and an antiferromagnetic superexchange  provides a new 
example 
of an "order from disorder" phenomenon ---  
an intermediate spin configuration 
(either a canted or a spiral state) is selected by  quantum and/or 
thermal 
fluctuations.  
\end{abstract}
] \narrowtext

Many magnetic systems of current experimental interest, for example
the colossal magnetoresistance (CMR) Manganites \cite{xener,anderson}
and Pyrochlores \cite{pyrochlores}, consist of itinerant electrons
interacting with an array of localized magnetic moments with spin $S$.
The simplest models of these systems comprise a single tight
binding band of electrons interacting with localized core spins by
a ferromagnetic (Hund's rule) exchange interaction $J_H \gg t$
\be
\label{KondoH}
{\mc H}_1 &=&
   - t \sum_{\langle ij \rangle\alpha} c^{\dagger}_{i\alpha}
c_{j\alpha}
   - J_H \sum_{i\alpha\beta}
\vec{S}_i . c^{\dagger}_{i\alpha}\vec{\sigma_{\alpha\beta}}c_{i\beta}
\en
where the sum $\langle ij \rangle$ is restricted to neighboring
sites.
At a classical level, the ground state of this model in $D>1$ must be
ferromagnetic, since the Hund's rule requires the spin of itinerant
electrons to be locally aligned with the core spins, and the
kinetic energy of electrons is in turn minimized by making all the
electron spins parallel.
This effect is usually called ``double exchange'',
and in this context the model of Eq. (\ref{KondoH}) is referred to 
as the double exchange ferromagnet (DEFM).

At a quasi--classical level the spin dynamics in a
DEFM can be described by an effective nearest--neighbor Heisenberg
model
$-J_1 \sum_{\langle ij \rangle} \vec{S}_i.\vec{S}_j$
with ferromagnetic exchange integral $J_1 = \overline{t}/4S^2$,
where $\overline{t}$ is the expectation value of the kinetic energy
per bond in the lattice 
\cite{degennes,kubo,furukawa95,plakida,millis,golosov}.
In addition, core spins also interact via a direct superexchange
\be
\label{exchange}
{\mc H}_2 &=& J_2 \sum_{\langle ij \rangle} \vec{S}_i.\vec{S}_j
\en
In CMR materials, $J_2$ is believed to be positive 
(antiferromagnetic),
and so competes with the DE mechanism.

A natural question to ask is how a system described by 
${\mc H} = {\mc H}_1 + {\mc H}_2$ evolves  towards an
antiferromagnet with increasing $J_2$ ?  As observed by de
Gennes \cite{degennes}, for classical spins ($S \rightarrow \infty$),
the FM becomes unstable at $J_1/J_2 =1$. At larger $J_2$, 
the competition between $J_2$ and the kinetic energy 
gives rise to an intermediate phase where 
 the neighboring lattice spins  are misaligned by an angle $\theta$,
where $\cos \theta/2 = J_1/J_2$ (in this regime, the
 DE model is no longer equivalent to 
Heisenberg model). 
However, $\theta$ alone does not specify a particular intermediate 
configuration ---
there exist an infinite set of classically degenerate
states with the same  $\theta$.  The two ends of this set are
the  two--sublattice canted phase
and the spiral phase~\cite{degennes}.

We performed a spin--wave analysis for the canted phase of the DEFM 
and found $\omega_{sw} (q) \equiv 0$ for all $q$.  This result 
reflects the local symmetry linking all the different classical
groundstates --- we can take any (classical) spin on the $A$ 
sublattice of the canted phase and rotate it about the direction of 
magnetization of the $B$ sublattice without changing the angle 
$\theta$.  
Since it costs no energy to make such an excitation, the system cannot
distinguish between different states, and is magnetically disordered 
even at $T=0$\cite{degennes}.
This argument, however, does not hold at finite $S$,
and we anticipate that quantum and/or thermal fluctuations
will enable the system to choose its true groundstate. Such ``order
from disorder'' effects have been widely discussed in the context of
magnetic insulators, \eg for Kagom\'e antiferromagnets
\cite{andrey} --- but not in the context of the metallic DEFM.

The most direct route to an answer to what kind of order is preferred
in the DE model is to determine the first instability of the DEFM, by
looking for the wave vector $Q^*$ at
which the spin--wave dispersion becomes unstable.
The quasi--classical arguments leading to an effective Heisenberg
model {\it cannot} answer this question since  they
predict a FM spinwave spectrum of the form
$\omega_{sw} (q) = 2 z S  (J_1 - J_2) (1 - \gamma_q)$,
where $z$ is the lattice coordination number, and
$\gamma_q = (1/z) \sum_\delta e^{i \vec{k} \vec{\delta}}$ where
$\{\vec{\delta}\}$ runs over nearest neighbors.
Thus $\omega_{sw} (q)$ vanishes identically for $J_2 =J_1$, and
no special wave vector is singled out.
In order to find the true groundstate for $J_2 > J_1$,
we then must go beyond the usual mapping of the DEFM onto an
effective Heisenberg model.

In  this paper we argue that the equivalence between DEFM and the
nearest--neighbor Heisenberg ferromagnet holds only for 
$S \rightarrow \infty$,
while fluctuation corrections in the DEFM are governed by fermions
and are different from those in the Heisenberg model.
We show analytically that for both small densities of electrons ($x 
\ll 1$)
and small densities of holes ($1-x \ll 1$), the first instability of 
the 
DEFM is against a two--sublattice canted structure with 
$Q^* = (\pi,\pi,\pi)$.
At intermediate densities ($0.3 \lesssim x \lesssim 0.9$ in $3D$),
the first instability is against a spiral spin configuration with
$Q^* \approx (0,0,0)$.  Similar results hold for the $2D$ case.
We also show that thermal corrections to $\omega_{sw}(q)$ compete 
with the
quantum corrections, and the trade--off between these two effects
gives rise to re--entrant transition between canted and spiral phases
with increasing $T$.  

The nonequivalence between DEFM and the Heisenberg ferromagnet has 
been
earlier detected in numerical studies~ \cite{kaplan,zang,t=0}.
Analytically, this has been demonstrated at $T=0$ by 
Golosov~\cite{golosov}.  
Our $T=0$ results agree with his, but we present more physical 
understanding. 
Our calculation scheme and results at $T \neq 0$ are entirely new.  

We now turn to the calculations.  Given that the the Hund's
rule coupling $J_H$ is the largest energy scale in the problem it is
desirable to diagonalize this term first and  project out all
electrons not locally aligned with the core spins.
We should also define spin--wave excitations such that they are
the true Goldstone modes of the order parameter, \ie transverse
fluctuations of the composite spin
$\vec{{\mc S}}_i = {\vec S}_i
+ \frac{1}{2}\sum_{\alpha\beta} c^{\dagger}_{i\alpha}
{\vec \sigma}_{\alpha \beta} c_{i\beta}$.
We can accomplish both of these goals by the following
procedure :
(i) introducing new Fermi operators $\f_i$ and $\p_i$ which
create electronic ``up'' and ``down'' states aligned with the
quantization axis of the composite spin, and (ii)
generalizing the Holstein--Primakoff
transformation to the case where the length of the spin is itself an
operator, introducing a corresponding bosonic operator $\at$. 
We work in the large $S$ limit usual for spin-wave theories.
Full details of the transformation will be given elsewhere 
\cite{usagain};
here we simply state that the transformation and its inverse 
are unitary, and satisfy all required (anti--)commutation relations 
\eg $\{\f,\fd \} = \{\p,\pd \} = [\at,\atd] = 1$, 
$[\f,\at]= \{\f,\p \} = 0$ \etc\      
We have also checked that our expansion procedure reproduces all 
features of 
an exact solution of the DE model on two sites \cite{shannon}.

\begin{figure}[tb]
\begin{center}
\epsfysize 6cm
\epsfxsize \columnwidth
\epsffile{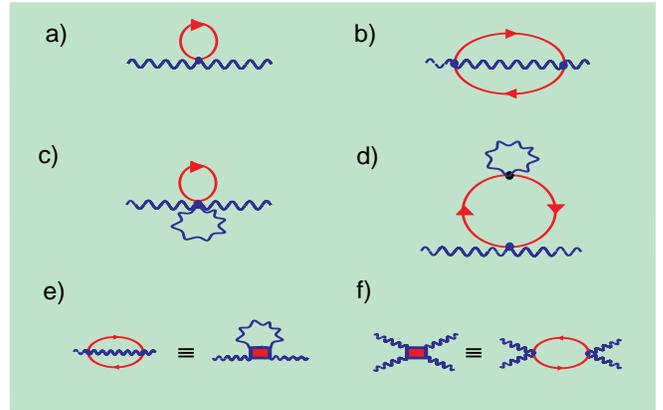}
\caption{a)--d)
Diagrams contributing to spin--wave self energy in Ferromagnet to
order ${\mc O}(1/S^2)$.  Only diagrams a) and b) are physically 
relevant.
e)--f) The representation of the diagram b) via an effective 
four--boson
interaction mediated by the charge suscpetibility 
(particle--hole polarization bubble).}
\label{fig:watermelon}
\end{center}
\end{figure}

Although the full analysis of the order from disorder phenomenon 
required us
to derive the transformation to order $1/S^2$, to capture the 
essential physics it is sufficient to know the inverse transformation 
to 
order $1/S$. To this order we have
\be
\label{inverse}
a &=& \at \left(1 + \frac{\pd\p - \fd\f }{4S}\right) -
\frac{\fd \p}{\sqrt{2S}} + {\mc O}(1/S^{3/2}) \no \\
\up &=& \f\left(1 - \frac{\atd\at + \pd \p}{4S}\right)
- \frac{\p \atd}{\sqrt{2S}} + {\mc O}(1/S^{3/2}) \no \\
\dn &=& \p\left(1 - \frac{\atd\at + \f \fd}{4S} \right)
+ \frac{\f \at}{\sqrt{2S}} + {\mc O}(1/S^{3/2})
\en
where $a$ is the Holstein--Primakoff boson associated with the 
core spin.
Substituting this transformation into (\ref{KondoH}), we find
that the Hund rule term in (\ref{KondoH}) does not contain any spin
operators and reduces to
\be
{\mc H}_1 &=& -\frac{J_H S}{2}
   \left[ \fd\f - \pd\p \left(1 + \frac{1}{S} \right)
   + \frac{\fd\f \pd\p}{S} \right]
\en
Clearly, $\p$
operators describe high--energy excitations and can be safely dropped.
Simultaneously, the hopping term in (\ref{KondoH})
transforms into the Hamiltionian which describes a single band of
spinless fermions interacting with (initially dispersionless)
Holstein--Primakoff bosons:
\be
\label{effectiveH}
{\mc H}_t &=&  \sum_{\vec{k}} (\epsilon_{k} -\mu) \fd_k \f_k +
\frac{1}{N} \sum_{\vec{k}_1\ldots \vec{k}_4}  V^{13}_{24}
 \fd_1 \f_2 \atd_3 \at_4 + \ldots
\en
 where $ \epsilon_{k} = -J_H S/2 -zt \gamma_k$, and
\be
\label{vertices}
\quad V^{13}_{24} &=&  \frac{zt}{4(S+\frac{1}{2})}
   \left[ (\gamma_1 + \gamma_2)
   \left( 1 + \frac{1}{8S} \right)
   - \left(\gamma_{1+3} + \gamma_{2+4}\right)
   \right]
\en
The perturbation theory for the
bosonic propagator $D (q, \Omega)$ is straightforward.
We have $D^{-1}(q,\Omega) = \Omega - \Sigma (q, \Omega)$.  The
self--energy physically comes from the fact that the oscillations  of
the core spins destroy perfect alignment of
electron spins, and this increases electron kinetic energy.
To ${\mc O}(1/S)$,  $\Sigma (q, \Omega)$ comes from a  single loop of
fermions Fig.~\ref{fig:watermelon}a and evaluates to
\be
\Sigma^{(1a)} (q) &=& 2 z J_1 S [1-\gamma_q]~;
J_1 = \frac{t}{4S^2} \frac{1}{N}
   \sum_{\vec{k}} n_k
   \gamma_k.
\en
so at this level we indeed reproduce the spin--wave
spectrum of a nearest--neighbor Heisenberg FM.

\begin{figure}[tb]
\begin{center}
\epsfysize 6cm
\epsfxsize \columnwidth
\epsffile{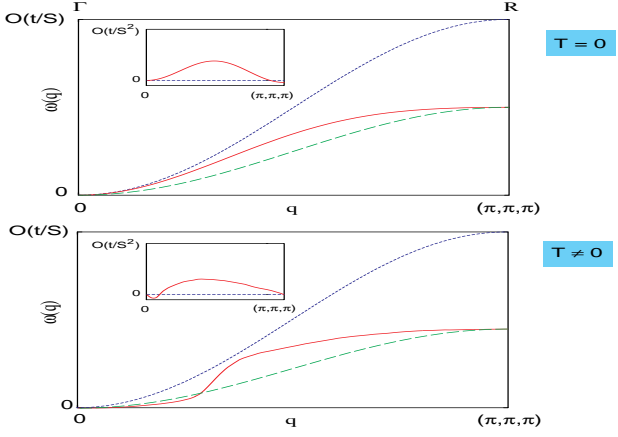}
\caption{Schematic corrections to the spin--wave dispersion 
in the $(1,1,1)$ direction
at zero and 
finite temperatures and small electron density $x$.   
Dotted lines show zero temperature classical (${\mc O}(1/S)$)
dispersion; dashed lines show renormalized classical dispersion
at ${\mc O}(1/S^2)$; solid lines show spectrum including all 
quantum/thermal 
effects at ${\mc O}(1/S^2)$.   Insets show residual dispersion 
at instability against canted ($T=0$) and spiral ($T\ne 0$) phases}
\label{fig:spinwave}
\end{center}
\end{figure}

The calculation of the spin--wave dispersion at higher order 
requires some care.  There are four self energy contributions at 
${\mc O}(1/S^2)$, one from each of the diagrams in 
Fig.~\ref{fig:watermelon}a--d.  
At finite temperature these include a contribution from a six--fold 
term in 
${\mc H}_t$, omitted in \protect{\mbox{Eqn. \ref{effectiveH}}}.   
However there is considerable cancellation between terms, and all 
important physical effects come from the diagram 
Fig.~\ref{fig:watermelon}b, which can be thought of as an effective
four--boson interaction mediated by Fermions 
(Fig.~\ref{fig:watermelon}e--f).
Assembling all contributions to the self--energy, splitting it
into quantum and thermal pieces, and neglecting the frequency
dependence of the self--energy (which is ${\mc O}(1/S^3)$ 
), we obtain a renormalized spin--wave dispersion
$\Omega (q) = 2 z J_1 S (1-\gamma_q) +
 \Sigma^{(2)}_{T=0}(q) +\Sigma^{(2)}_T (q)$
where
\be
\Sigma^{(2)}_{T=0}(q)
    &=& -\frac{zt}{4S^2} \frac{1}{N^2}
            \sum_{p, l} n_F (p) (1 - n_F (l)) \no \\
&&
   \times \left[(1 -\gamma_q) \gamma_p
        \frac{\gamma_p + \gamma_l}{\gamma_p - \gamma_l} -
\frac{\gamma^2_p -
        \gamma^2_{q+p}}{\gamma_p - \gamma_l}
    \right]
\label{T=0}
\en
and
\be
&&\Sigma^{(2)}_T (q) = -\frac{zt}{4S^2} \frac{1}{N^2}
\sum_{p, l } n_B (l) \frac{n_F (p + \frac{m}{2}) -
n_F (p - \frac{m}{2})}{\gamma_{p+m/2} - \gamma_{p-m/2}} \no \\
&& \quad \times (\gamma_{p-m/2} - \gamma_{p+q -m/2})(\gamma_{p+m/2} -
\gamma_{p+q-m/2}).
\label{T}
\en
Here $ {\vec m} = {\vec q}-{\vec l}$,
and $n_B (q) = n_B (\Omega (q))$ and $n_F (k) = n_F (\epsilon_k)$
are Bose and Fermi distribution functions, respectively.
The latter can be approximated by step functions as we will be
interested in $T < J_1S \ll t$.

We now analyze the form of $\Sigma^{(2)} (q)$, beginning
with the case $T=0$.  The first term in (\ref{T=0})
simply renormalizes the classical Heisenberg--like spinwave 
dispersion, 
while the second has a dependence on $q$ which is quite different 
from that in the nearest--neighbor Heisenberg model.  
This term is either positive or negative
throughout the Brillouin zone, depending on the electronic density, 
and
is symmetric under
${\vec q} \rightarrow {\vec \pi} - {\vec q}$
where
${\vec \pi} = (\pi,\pi,\pi)$.
Along the zone diagonal, it reduces to
$\frac{zt}{4S^2}I(x) \left[ 1 -\cos(2q) \right]$
where
$I(x) = (1/N^2) \sum_{p l} n_F (p) (1 - n_F (l))
~(\gamma_p^2 - \gamma_{p+\pi/2}^2)/(\gamma_p - \gamma_l)$
changes sign twice as a function of $x$.
This form of the correction to Heisenberg dispersion
is comparable to that found 
in numerical studies of the DE model on a ring \cite{kaplan}.

We find that for intermediate densities
$0.31(7) < x < 0.92(7)$ in $2D$, and $0.31(7) < x < 0.94(2)$ in $3D$,
$I(x) < 0$ and quantum effects cause a relative softening of spinwave 
modes near the zone center.  This means that the first instability 
of the DEFM with competing AF exchange interactions
is against a spiral phase with $Q^* \approx (0,0,0)$.
On the other hand, for a small density of electrons (or holes), 
$I(x) >0$, and quantum effects 
instead lead to a relative softening of modes near the zone boundary.
In this case the spin--wave spectrum first becomes
unstable against a canted spin configuration with 
$Q^* = (\pi,\pi,\pi)$. 
These results are in perfect agreement with earlier studies~\cite{t=0}. 
Finally, we note that if we formally extend our 
large $S$ analysis to arbitrarily small $S$, there exists
a critical value of spin $S = S^* \approx 1$ (in 3D) 
for which the DEFM becomes 
unstable {\it even in the absence of AF interactions}.
This opens up the possibility that for small $S$ (\eg, $S=1/2$), 
the ground state of the DE model may not be a FM, as suggested by some
numerical studies~\cite{zang}.   

We now proceed to finite $T$.  In a Heisenberg model, finite 
temperature effects do not change the form of spinwave dispersion,
but the overall scale of the dispersion is reduced by a factor 
$\propto T^{5/2}/S^2$ \cite{sw}.
In the case of the DEFM, the result is more complex and coincides 
with the behavior of a Heisenberg FM only in a very limited range of 
frequencies and temperatures. 
Indeed, there are three  typical momenta in our problem
--- an external $q$, a fermionic $p_F$, and a typical bosonic 
momentum $l_{typ} \sim (T/J_1S)^{1/2}$ 
(in units where the lattice constant $a=1$). 
Consider for definiteness the case of small $x$ when 
$p_F = (6\pi^2 x)^{1/3} \ll 1$.
Evaluating the self--energy (\ref{T}) by expanding all $\gamma_k$
factors to second power in momenta, we obtain in 3D, for 
$q \ll p_F,(T/J_1S)^{1/2}$
\be
\Sigma^{(2)}_T (q)  = - \frac{t q^2 p_F}{144 \pi^2 S^2} \frac{1}{N}
\sum_l n_B (l) l^2 \Phi \left(\frac{l}{2\p_F}\right)
\label{t1}
\en
where $\Phi(\nu)$ is a smooth function with
$\Phi (\nu \rightarrow 0) =1,~\Phi (\nu=1) =0$,
$\Phi(\nu \gg 1) \approx -1/\nu^2$,
and
$(1/N)\sum_l n_B(l) l^2 \approx (3 \zeta(5/2)/16 
\pi^{3/2})~(T/J_1S)^{5/2}$.
We see that $\Sigma^{(2)}_T (q)$
agrees with the Heisenberg model only at
$(T/J_1S)^{1/2} \leq p_F$. In the opposite limit,
we have $\Sigma^{(2)}_T (q) \propto q^2 p^3_F (T/J_1S)^{3/2}$.
Moreover, the sign of $\Sigma^{(2)}_T (q)$ changes between the two
limits.

For large external $q$: $q \gg p_F, (T/J_1S)^{1/2}$, we obtain
\be
\Sigma^{(2)}_T (q)  =  \frac{t~ p^3_F}{24 \pi^2 S^2 N}
\sum_l n_B (l) l^2 \left[1+\gamma_q - \frac{1}{3}
\frac{1-\gamma_{2q}}{1-\gamma_q}\right]
\label{t2}
\en
We see that at these $q$, the temperature correction to the spin--wave
energy is positive, and is of order
$p^3_F (T/J_1S)^{5/2}$. It depends only
weakly on $q$ at intermediate momenta, and vanishes at
$q=(\pi,\pi,\pi)$. This result is indeed very different from that in
the Heisenberg FM. The temperature correction to the spin-wave 
dispersion 
is shown schematically in Fig.\ref{fig:spinwave}a.
A very similar behavior holds in 2D case, the only difference is that
the $2D$ analog of the scaling function $\Phi(\nu)$ does not change 
sign at
$\nu \sim {\mc O}(1)$ but rather behaves as $1/\nu^4$ at large $\nu$.

\begin{figure}[tb]
\begin{center}
\leavevmode
\epsfysize 40mm
\epsffile{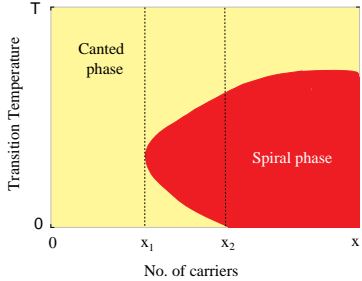}
\caption{Phase diagram showing first instability as a function
of carrier concentration $x$ and temperature $T$.  For some
range of dopings $x_1 < x < x_2$ a re--entrant phase transition 
is predicted.
}
\label{fig:phasediagram}
\end{center}
\end{figure}

The different nature of quantum/thermal effects in the DEFM 
and the Heisenberg FM is a result of the different nature of the 
coupling between spins.
In a Heisenberg FM, the $q^2 T^{5/2}$ form of the temperature
correction 
comes from the fact 
that four--boson vertex scales as 
$q^2 l^2$~\cite{sw}.  
In the DEFM, the interaction between spin waves is mediated by 
the charge susceptibility of the Fermi gas,
$\Pi (q-l)$ (see Fig. (\ref{fig:watermelon}e)).
At $q,l_{typ} \ll p_F$, $\Pi (q-l)$ can be approximated by
$\Pi (0)$, and the effective interaction has the same form as in the
Heisenberg FM. 
However, when $|q -l| \gg p_F$, the susceptibility decreases as
$\Pi (m) \propto p^2_F/m^2$. This replaces either one power of
$T/J_1S$ or $q^2$ factor by $p^2_F$ in the renormalization of the 
dispersion, exactly as we found.

The unusual temperature dependence of $\Sigma^{(2)}_T (q)$ gives 
rise to the possibility of a re--entrant transition between spiral 
and canted states with varying $T$.
As discussed above, at small $p_F$, quantum fluctuations favor
canted phase. On the other hand, classical fluctuations at low
$(T/J_1S)^{1/2} < p_F$
soften the dispersion near $q=0$ and hence favor the spiral state. As
a result, the transition line between the two states bends towards
smaller $x$ at finite $T$.  However, as $T$ increases and becomes
larger than $J_1S p^2_F$, the sign of $\Sigma^{(2)}_T (q)$ changes,
and thermal fluctuations now
favor the canted phase. As a result, the transition line now
bends  towards higher $x$ with increasing $T$. This
can give rise to a re--entrant transition --- when $T$
increases at at a given, small $x$, the canted state first becomes
unstable towards the spiral state, and then returns back at even
larger $T$.  A possible phase diagram for $x \ll 1$ is shown schematically 
in Fig.\ref{fig:spinwave}b.
The region occupied by the canted phase first shrinks and then 
expands with increasing $T$.  
Qualitatively similar behavior will 
hold for $1-x \ll 1$.

To summarize, in this paper we introduce a novel large $S$ expansion scheme
for systems with strong Hund's rule coupeling, and use it to show 
how and why the spin dynamics of a DEFM differ from those of a Heisenberg FM.
We find that the two models are equivalent at the classical level 
($S \rightarrow \infty$), but that both quantum and thermal
corrections in the DEFM are different because the interaction between 
spinwaves is mediated by fermions.
We also find that the DEFM  provides a new example of an "order
from disorder" phenomenon --- in the case of a competition between
ferromagnetic double exchange and a direct antiferromagnetic
superexchange,
the classical ground state is infinitely degenerate, but fluctuations
lift the degeneracy and select a true intermediate spin
configuration, either
a canted state or a spiral state, depending on the electron density.
This together with the calculated temperature corrections
gives rise to the possibility of an unusual re--entrant transition 
between spiral and canted states with varying $T$.

The softening of zone boundary spin--waves which we predict 
{\it has} been observed in neutron
scattering experiments on the CMR Manganites \cite{neutrons}, but
at a {\it hole} doping $\tilde{x} = 1-x \sim 0.3$ for which our  zero
temperature theory would predict a relative hardening of the 
dispersion at $(\pi,\pi,\pi)$. 
It has been suggested that this softening may be due to
the influence of optical phonons \cite{furukawa99}, or orbital
degrees of freedom \cite{khaliullin}.    We note, however, that the 
DE mechanism {\it does} predict a softening at the physically 
relevant doping, for high enough temperatures.
A careful experimental examination of the temperature dependence of 
the spin--wave spectrum is therefore necessary to determine which 
mechanism contains the relevant physics.

The issue left for further studies
 is a possible phase separation in the non-ferromagnetic 
regime~\cite{elbio}. 
To study this possibility in our approach, one 
has to analyze the sign of the 
longitudinal susceptibility in, e.g., canted phase.  If it is 
negative, then the system is unstable towards phase 
separation~\cite{ps}.
There calculations are  currently under way.

It is our pleasure to acknowledge helpful conversations with J.\
Betouras, D.\ Golosov, R.\ Joynt and M. Rzchowski.
This work was supported under NSF grant DMR--9632527.

\end{document}